\documentclass[conference,10pt]{IEEEtran}
\usepackage{graphicx,amsmath,times,setspace,bm}
\IEEEoverridecommandlockouts
\usepackage{cite}
\usepackage{array}
\usepackage{subfig}
\usepackage[ruled,vlined]{algorithm2e}
\usepackage[T1]{fontenc}
\usepackage[mathscr]{eucal}
\usepackage{mathptmx}
\usepackage{multirow}
\DontPrintSemicolon

\usepackage{balance}

\let\mathcal\relax
\DeclareMathAlphabet\mathcal{OMS} {cmsy}{b}{n}
\SetMathAlphabet \mathcal{normal}{OMS}{cmsy}{m}{n}
\DeclareMathAlphabet\mathbcal{OMS} {cmsy}{b}{n}

\DeclareMathOperator*{\argmin}{\arg\!\min}

\newcommand{\cP}{\mathcal{P}}

\newcommand{\equals}{=}

\newcommand{\plus}{\!+\!}

\newcommand{\dB}{\, \mathrm{dB}}

\begin{document}

\title{Cooperative Relaying in LoRa Sensor Networks \vspace*{-12mm}\\
{\footnotesize \textsuperscript{}}
\thanks{This work has been supported by the K-project DeSSnet (Dependable, secure and time-aware sensor networks), which is funded within the context of COMET -- Competence Centers for Excellent Technologies by the Austrian Ministry for Transport, Innovation and Technology (BMVIT), the Federal Ministry for Digital and Economic Affairs (BMDW), and the federal states of Styria and Carinthia; the COMET program is conducted by the Austrian Research Promotion Agency~(FFG).}
}

\author{\IEEEauthorblockN{Siddhartha S. Borkotoky\textsuperscript{1}, Udo Schilcher\textsuperscript{1,2}, and Christian Bettstetter\textsuperscript{1,2}}
\IEEEauthorblockA{\textit{\textsuperscript{1}Lakeside Labs GmbH, \textsuperscript{2}Institute of Networked and Embedded
Systems, University of Klagenfurt} \\
Lakeside Science \& Technology Park, 9020 Klagenfurt am W\"{o}rthersee, Austria \\
borkotoky@lakeside-labs.com}\vspace{-0.8cm}
}

\maketitle

\begin{abstract}
We propose a communication scheme with relays to improve the reliability of a Long Range (LoRa) sensor network with duty-cycle limitations. The relays overhear the sensors' transmissions and forward them to a gateway. Simulations show that relaying is very beneficial, even though the nodes are not coordinated and duty cycling limits the number of sensor measurements that can be forwarded. In our setup, a single relay can halve the measurement loss rate and eight relays provide a gain of up to two orders of magnitude. Further improvements are achieved by including a few past measurements in each~frame.
\end{abstract}
\begin{IEEEkeywords}
LoRa, relay, sensor network, reliability
\end{IEEEkeywords}

\renewcommand{\baselinestretch}{0.9}
\small\normalsize

\section{Introduction}
\label{intro}
We study an industrial network with sensors periodically capturing physical parameters and sending their readings to a control station via a gateway. This work is motivated by specific use cases pertaining to predictive maintenance in a chemical production plant, where sensors installed over a small region\,---\,a production floor,  warehouse, or similar environment\,---\,transmit small amounts of data every few seconds. The gateway is in a separate room and connected to a control station. The environment is populated with metallic structures and moving objects.

The LoRa wireless technology~\cite{BRV16} is suited for applications of this kind due to its robustness against multipath effects. Nevertheless, frame losses are still expected due to time-varying propagation loss~\cite{MRP17} and interference~\cite{GeR17}. Since LoRa is commonly deployed in unlicensed frequency bands~\cite{BRV16}, it is often subject to regulations that limit a transmitter's duty cycle (e.g., to $1\,\%$ in the 868~MHz band in the European Union~\cite{AVT17}). One implication of this duty-cycle constraint is that the gateway could send only a small number of acknowledgements in response to frames received from the sensors. Acknowledgement-based retransmission protocols for improved reliability are thus ineffective for systems with many sensors.

\begin{figure}
\centering
\includegraphics[scale=0.27,bb=325 0 520 520]{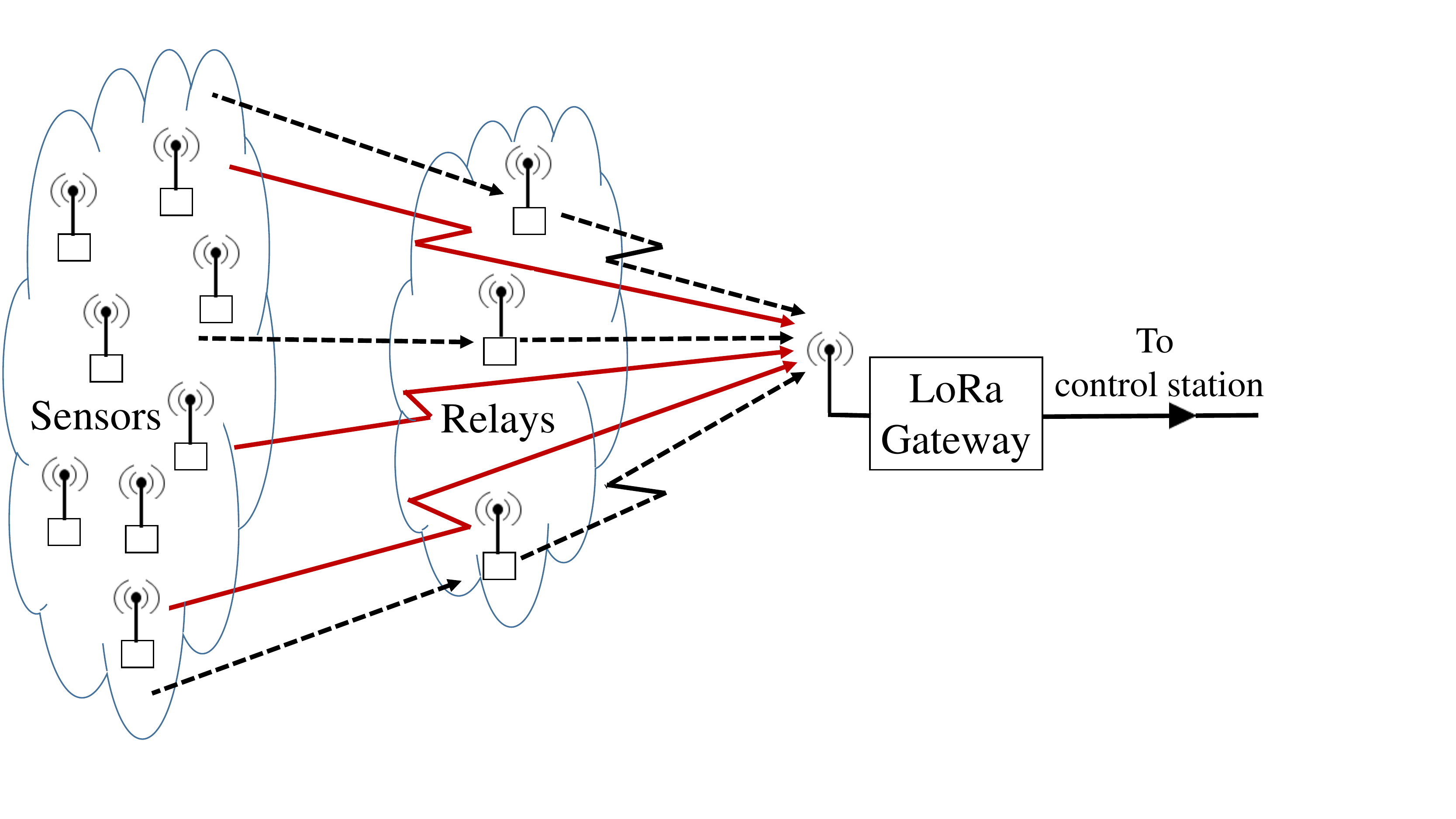}
\setlength{\belowcaptionskip}{-12pt}
\caption{Diagram of a LoRa sensor network with relays.}
\label{net_diagram}
\end{figure}

We therefore investigated the use of repetition redundancy \cite{BBS19} and found that the number of measurement losses can be significantly reduced by including a few recent past measurements in each frame. The paper at hand extends this work: in addition to employing repetition redundancy, we place decode-and-forward relays between the sensors and the gateway to overhear transmissions from the sensors and forward overheard measurements to the gateway (see Fig.~\ref{net_diagram}). The gateway is able to obtain a measurement if it receives at least one frame containing the measurement, either directly from the sensor or via at least one of the relays. The sensors are battery powered. The relays are connected to power supplies (as in \cite{DiG19}) since they can be placed at convenient locations in the building.

\renewcommand{\baselinestretch}{0.9}
\small\normalsize

Our objective is to examine the reliability improvement provided by the relay nodes. Of course, the use of relays will reduce the number of measurements lost, but it is unclear whether this improvement will be substantial or not\,---\,in a setup with  some relays (up to 16) and many sensors (more than 100), where all nodes are limited in their capabilities due to duty cycling and other constraints~\cite{AVT17}.

The relaying scheme we use entails low complexity and can be implemented with off-the-shelf LoRa devices. Sensors do not need to be aware of the relays' presence. Nodes neither synchronize nor exchange control information or acknowledgements. Relays do not know if measurements are forwarded by other relays. Since a relay must adhere to the duty-cycle limit, it may not be able to forward all overheard measurements. 

Our simulation results demonstrate that\,---\,despite the aforementioned limitations\,---\,even a few relays significantly improve the probability of a sensor's measurement reaching the gateway.
The paper is organized as follows: Section~\ref{related_work} provides an overview of prior work. Section~\ref{LoRa_description} addresses relevant features of LoRA. Section~\ref{relay_protocol} describes the relaying scheme. Section~\ref{reliability_analysis} derives an expression for the probability of a gateway failing to receive a measurement. Section~\ref{perf_res} presents and discusses the performance by means of simulations. Section~\ref{conclusion} concludes the paper.

\section{Related Work} \label{related_work}
A multihop communication scheme for LoRa networks is described in~\cite{DiG19}, where a distance-vector routing protocol is employed to construct routes between sensors and gateway. A cooperative communication strategy for multihop LoRa networks is presented in~\cite{LZK17}, where concurrent transmissions by the relays are employed. Concurrent transmissions in a multihop LoRa network are also utilized in~\cite{BVR16}. The differences to our work are as follows: The methods of~\cite{DiG19}--\hspace{1sp}\cite{BVR16} all depend on time synchronization among devices. We focus on a simpler implementation, where the relays play an auxiliary role and are invisible to the sensors, and synchronization among the sensors themselves or between sensors and relays is not required. No routing algorithms are used; instead, a relay tries to overhear frames from any transmitting sensor. Unlike our work, \cite{DiG19}--\hspace{1sp}\cite{BVR16} do not provide an analytical characterization of the performance. A difference from other work on cooperative relaying (not specific to LoRa) is that LoRa's proprietary transceiver design does not permit all forms of diversity combining, such as maximal-ratio combining~\cite{CTS15}.

The work \cite{MRP17} considers the transmission of redundancy in LoRa. The use of relays, the presence of multiple transmitters, and the resulting interference are not investigated. The redundancy is generated via application-layer coding on past data. To avoid raising the computational burden, we do not consider coding at the sensors. Coding at the relays is not feasible since a large overhead must be sent in the form of encoding vectors, and duty-cycle constraints preclude the transmission of such~overhead.



\section{LoRa Features} \label{LoRa_description}
A technical introduction to LoRa can be found in~\cite{Sem13} and~\cite{Sem15}. Let us summarize the features relevant to the discussion in this paper.
The LoRa physical layer employs chirp spread spectrum~\cite{Sem15}. The {spreading factor} $s$, which takes integer values between 7 and 12, determines the ratio between the symbol rate and the chip rate of the signal. A higher $s$ provides greater immunity to thermal noise at the expense of a longer frame duration. The symbol duration is \mbox{$t_{\mathrm{sym}}(s) = 2^{s}/w$}, where $w$ is the signal bandwidth. The payload in a frame is preceded by a preamble for synchronization and an optional header. The duration of a frame with $b$ bytes of payload is $t_{\mathrm{fr}}(b,s) = t_{\mathrm{pr}}(s) + t_{\mathrm{pl}}(b,s)$,
where $t_{\mathrm{pl}}(b,s)$ is the duration of the payload and $t_{\mathrm{pr}}(s)$  is the preamble duration, given by $t_{\mathrm{pr}}(s) = (n_{\mathrm{pr}} + 4.25)\: t_{\mathrm{sym}}(s)$,
where $n_{\mathrm{pr}}$ is the number of preamble symbols in the frame~\cite{Sem13}. The payload duration is $t_{\mathrm{pl}}(b,s)
=\left[8 + \max\left\{ \left\lceil \frac{2b-s-5h+11}{s-2l} \right\rceil (c+4), 0\right\}\right]\,t_{\mathrm{sym}}(s)$,
where $h=1$ if a header is included and 0 otherwise, $l=1$ if low data rate optimization is enabled and 0 otherwise, and $c$ depends on the rate of the channel code employed for the frame and can take integer values between 1 and~4~\cite{Sem13}.

Three classes of operation are defined for end devices~\cite{AVT17}: Class~A employs unslotted ALOHA for uplink transmissions; after a transmission, a device monitors any response from the gateway during two receive windows. Class~B allows for additional receive windows scheduled by the gateway via synchronizing beacons. Class~C devices remain in the receive mode while not transmitting.

LoRa waveforms with different spreading factors are orthogonal to one another. Hence, for two frames to interfere, they must employ the same spreading factor and overlap in frequency. LoRa waveforms exhibit the capture effect: When simultaneously received signals of differing power use the same channel and spreading factor, the strongest signal is correctly demodulated and decoded, whereas the frames carried by the weaker signals are lost. Field measurements show that LoRa frames are correctly received if the strongest of the interfering signals is at least $6\,\dB$ weaker than the desired signal~\cite{BRV16}.

\section{The Relaying Scheme} \label{relay_protocol}
We consider $n$ nodes sensing physical parameters every $t$ seconds and employing Class A LoRa communications to send their measurements to a gateway. All sensors have the same transmit power and are not synchronized. In addition to sending the current measurement in each transmission, a sensor has the option to employ repetition redundancy by adding the $r$ most recent measurements to the frame. Suppose that $\omega$ relays \mbox{($\mathrm{r}_1$, $\mathrm{r}_2$, \ldots, $\mathrm{r}_{\omega}$)} are placed between the sensors and gateway. The relays operate in Class~C, which means they are in the receive mode except when transmitting. All relays are powered by external supplies to enable continued reception (as in~\cite{DiG19}). Each relay periodically switches between overhearing the sensors' transmissions during a receive window of $t_\mathrm{rx}$ seconds and forwarding the contents of overheard frames to the gateway during a transmit window of $t_\mathrm{tx}$ seconds, where $t_\mathrm{tx}/(t_\mathrm{rx} \plus t_\mathrm{tx})$ is no greater than the maximum permitted duty cycle. A relay stores the current measurement of each overheard frame in its buffer; past measurements, if any, are not stored. The measurements collected during a receive window are placed in a frame and\,---\,along with the identifier (ID) of the sensor that produced it\,---\,sent to the gateway during the subsequent transmit window.  Note that the duration of a relay's frame cannot exceed $t_\mathrm{tx}$ seconds. If a relay receives in the preceding receive window more measurements than fit in a frame of maximum duration, it (randomly) discards some measurements. Due to their proximity to the gateway, the relays employ a lower spreading factor than the sensors; hence, their transmissions do not interfere with those from the sensors. The relays transmit over orthogonal time slots; the slot assignment is straightforward for a small number of relays.

\section{Measurement Loss Probability} \label{reliability_analysis}
The gateway may receive a measurement directly from the sensor or via one or more relays. The \emph{measurement loss probability} (MLP) is the probability of a measurement failing to reach the gateway via any of these paths:
\begin{eqnarray}
\mathrm{MLP}(r) &=& P_{\mathrm{dir}}(r)\,\cdot  \prod_{i=1}^{\omega} P_{\mathrm{r}_i}(r)\:.
\end{eqnarray}
The term $P_{\mathrm{dir}}$ is the probability that the gateway fails to receive a measurement directly from the sensor. The term $P_{\mathrm{r}_i}$ is the probability that the gateway fails to receive a measurement via relay $\mathrm{r}_i$, which is given by
\begin{equation}\nonumber
P_{\mathrm{r}_i}(r) = 1 - P_{{\text{rw,r}_i}}(r)(1\!-\!P_{{\text{s-r}_i}}(r))(1\!-\!P_\mathrm{drop, r_i}(r)) (1\!-\!P_{\mathrm{r}_i\text{-}\mathrm{g}})\,,
\end{equation}
where $P_{{\text{rw,r}_i}}$ is the probability that the first frame containing the measurement is transmitted in the receive window of relay $\mathrm{r}_i$, $P_{{\text{s-r}_i}}$ is the probability that relay $\mathrm{r}_i$ fails to receive a frame transmitted by a sensor, $P_\mathrm{drop, r_i}$ is the probability that $\mathrm{r}_i$ discards a received measurement due to frame duration exceeding $t_{\mathrm{tx}}$ seconds, and $P_{\mathrm{r}_i\text{-}\mathrm{g}}$ is the probability that the gateway fails to receive a frame sent by $\mathrm{r}_i$.

The spreading factors used by sensors and relays are $s_\mathrm{sen}$ and $s_\mathrm{rel}$, respectively. The duty cycle of a sensor~is
\begin{equation}
f(r,s_{\mathrm{sen}}) = \frac{t_{\mathrm{fr}}((r+1)\beta,s_{\mathrm{sen}})}{t}\:,
\end{equation}
where $\beta$ is the number of bytes required to represent one measurement. We assume that the distances $D$ from the sensors to the gateway are independent and identically distributed, as are the fading gains $A$ on the links. With transmit power $\cP$, signal wavelength $\lambda$, and pathloss exponent $\alpha$, we use $\gamma = (\lambda/4\pi)^{\alpha}\cP$. We assume that a transmitter selects one of $n_c$ channels uniformly at random.

The probability that the gateway receives none of the \mbox{$r \plus 1$} sensor transmissions containing a certain measurement~is
\begin{equation} \label{p_dir}
P_{\mathrm{dir}}(r) = (1 - (1-P_i(r))(1-P_f(r)))^{r+1}\:,
\end{equation}
where $P_i$ and $P_f$ are the probabilities of an outage due to interference and fading, respectively, for a transmission from a sensor to the gateway. It follows from~\cite{BBS19} that
\begin{eqnarray} \nonumber \label{S_I_general}
P_i(r) &=& \int_{S_D} \int_{S_A} (1 - F_M (0.25 \gamma a w^{-\alpha})) f_A(a)f_D(w)\:\mathrm{d}a\:\mathrm{d}w \\
&=& 1 - \int\displaylimits_{S_D} \int\displaylimits_{S_A}  e^{-\kappa(a,w,r)} f_A(a)f_D(w)\:\mathrm{d}a\:\mathrm{d}w\:,
\end{eqnarray}
where $S_D$ and $S_A$ are the supports of $f_D(\cdot)$ and $f_A(\cdot)$, and
\begin{align}
\kappa(a&,w,r) = \\ \nonumber
&n_c^{-1} (n - 1)  f(r, s_{\mathrm{sen}})[1- \int_{S_D} F_A(0.25 a u^{\alpha} w^{-\alpha}) f_D(u)\:\mathrm{d}u]. \nonumber
\end{align}
The probability of an outage due to fading is~\cite{BBS19}
\begin{eqnarray} \label{S_f_general} \nonumber
P_f(r) &=& \int_{S_D} \mathrm{P}(\gamma A u^{-\alpha} < \psi)\, f_D(u) \:\mathrm{d}u\\
&=& \int_{S_D} F_A(\gamma^{-1} u^{\alpha} \psi)\,f_D(u)\:\mathrm{d}u\:.
\end{eqnarray}

With appropriate substitution of parameters, the expressions above can be used to determine $P_{{\text{s-r}_i}}$ and $P_{\mathrm{r}_i\text{-}\mathrm{g}}$  as well. Since the relays transmit in orthogonal time slots, only outages due to fading need to be considered for $P_{\mathrm{r}_i\text{-}\mathrm{g}}$; in the absence of this orthogonality assumption, interference among the relays would have to be accounted for, too.

The probabilities $P_{{\text{rw,r}_i}}$ and $P_\mathrm{drop, r_i}$ cannot be obtained from the analysis in~\cite{BBS19}. We derive their expressions as follows. The probability that a transmission is made within relay $\mathrm{r}_i$'s receive window is
\begin{equation}
P_{{\text{rw,r}_i}}(r) = \frac{t_\mathrm{rx} - t_f(r)}{t_\mathrm{rx} + t_\mathrm{tx}},
\end{equation}
where $t_f(r) = t_{\mathrm{fr}}((r+1)\beta,s_{\mathrm{sen}})$ is the duration of a sensor's frame. This follows from the observation that for a complete frame to be received during a relay's receive window, the transmission must start within the first $t_\mathrm{rx} - t_f(r)$ seconds of a \mbox{$(t_\mathrm{rx}+t_\mathrm{tx})$-second} period that includes a receive window followed by a transmit window. Due to the lack of synchronization, the transmission start is treated as a uniform random~variable.

\begin{figure}
\centering
\includegraphics[scale=0.3,bb=100 270 520 550]{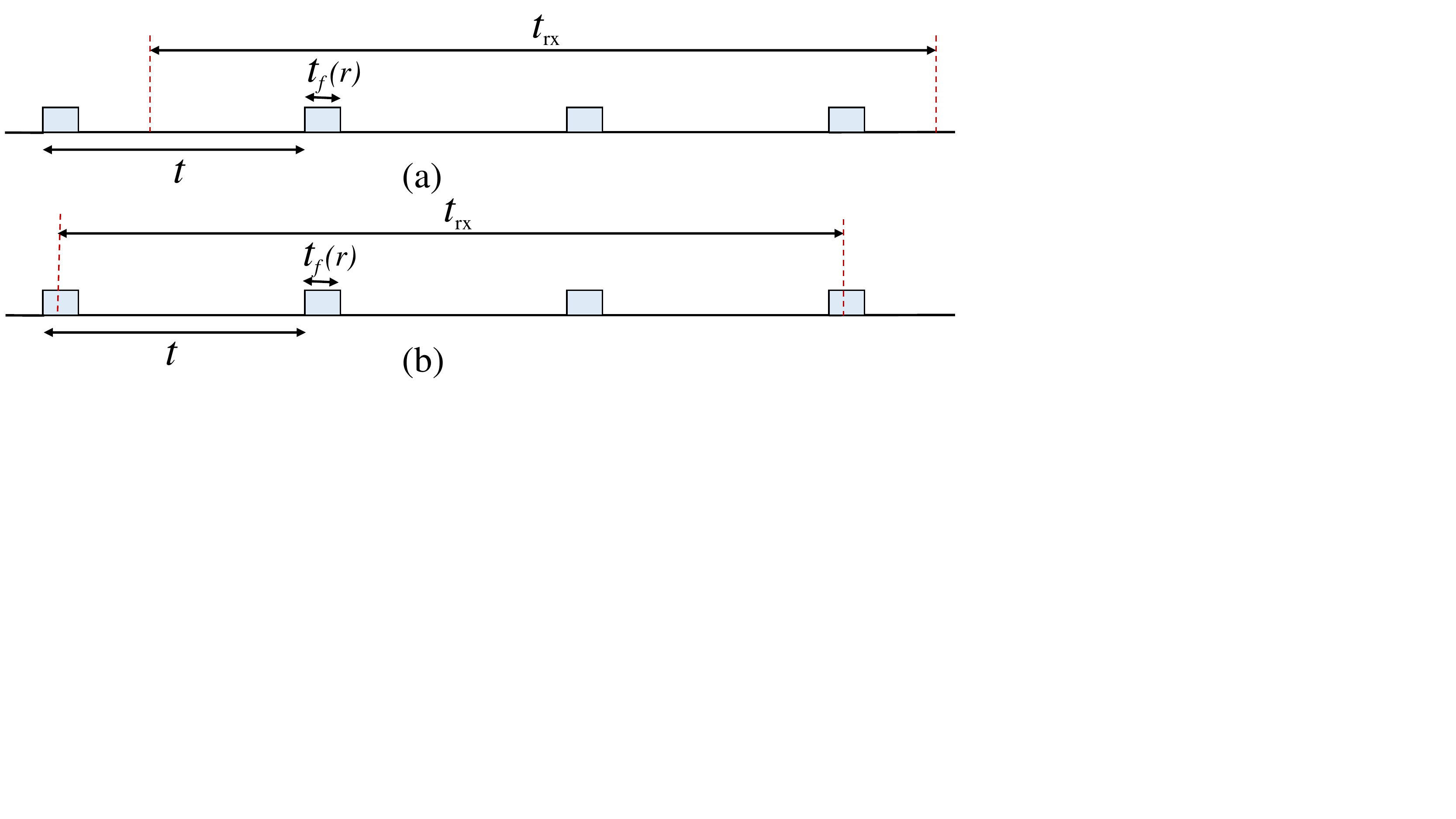}
\setlength{\belowcaptionskip}{-37pt}
\caption{A relay's receive window vis-\`{a}-vis a sensor's transmit windows ($\xi = 3$).}
\label{timing_diag}
\end{figure}

To determine the probability $P_\mathrm{drop, r_i}$ that relay $\mathrm{r}_i$ drops a measurement due to lack of space in the frame, let $Y$ be the total number of frames sent by the sensors during the receive window of a relay and $Z$ be the number of frames the relay received. For simplicity, suppose $t_\mathrm{rx}$ is an integer multiple of $t$ and define $\xi = t_\mathrm{rx}/t$.  To determine the probability mass function (pmf) of $Y$, note that if the start of a relay's receive window is outside a sensor's transmit window (the duration of $t_f(r)$ seconds, over which a sensor transmits a frame), the sensor transmits $\xi$ complete frames during the relay's receive window; otherwise, $\xi - 1$ complete frames are transmitted. Examples of the two situations  for $\xi = 3$ are illustrated in Fig.~\ref{timing_diag}, in which (a) and (b) show the transmission of three and two complete frames over $t_\mathrm{rx}$ seconds, respectively. The probability that the start of a relay's receive window is outside of a sensor's transmit window is $p = 1 - t_f(r)/t$. Therefore, $Y = \mu + \eta$, where $\mu = n(\xi - 1)$ and $\eta$ is binomial $(n,p)$.

\renewcommand{\baselinestretch}{0.9}
\small\normalsize

Let $v$ be the maximum number of measurements allowed to be included in a frame sent by a relay. Thus,
\begin{eqnarray}
v &=& \max \{ n : t_{\mathrm{fr}}(n(\beta + l),s_{\mathrm{rel}}) \leq t_{\mathrm{tx}} \},
\end{eqnarray}
where $l$ is the length of a sensor ID in bytes. Clearly, $P_\mathrm{drop, r_i}(r) = 0$ for $v \geq \mu + n$. For $v < \mu + n$, conditioned on $Z = z$, the probability that a measurement is dropped is $\max\{{z - v, 0\}} / z$. Conditioned on $Y = y$, $Z$ is binomial $(y,\theta)$, where $\theta = P_{{\text{s-r}_i}}(r)$.  Therefore, we have
\begin{eqnarray} \nonumber \label{drop_probability}
P_\mathrm{drop, r_i}(r)\!\!\!\! &=& \!\!\!\! \sum_{y = \mu}^{\mu + n}\sum_{z = 0}^{y}\frac{\max\{z-v,0\}}{z} \mathrm{P}(Y=y,Z=z) \\  \nonumber
&=& \!\!\!\!\sum_{y = \zeta}^{\mu + n}\sum_{z = v+1}^{y}\left(1\!-\!\frac{v}{z} \right)\mathrm{P}(Z=z|Y=y)\mathrm{P}(Y=y) \\ \nonumber
&=& \!\!\!\!\displaystyle\sum_{y = \zeta}^{\mu+n}\sum_{z = v+1}^{y}\left(1\!-\!\frac{v}{z} \right) {y \choose z}(1-\theta)^z {\theta}^{y-z}  \\
&& \!\!\!\!\!\!\!\!\hspace*{15mm} {n \choose y-\mu} (1-p)^{n-y+\mu} p^{y-\mu}\:,
\end{eqnarray}
where  $\zeta\!=\!\max \{\mu, v\!+\!1\}$.  Ignoring the possibility of incomplete frames, the following approximation is~obtained:
\begin{align}  \label{p_drop_p_zero}
P_\mathrm{drop, r_i}(r) &\approx \displaystyle\sum_{z = v+1}^{\mu + n}\left(1 - \frac{v}{z} \right) {\mu+n \choose z}(1-\theta)^z {\theta}^{\mu+n-z}.
\end{align}

\section{Performance Analysis}
\label{perf_res}
\subsection{Simulation setup}

We extend LoRaSim~\cite{LoRaSim} to support relays, Nakagami fading, and periodic transmissions, and use it to simulate the transmission of measurements from $n$ sensors to a gateway in a frequency band with a duty-cycle limit of $1\,\%$. The nodes are simulated as points on a two-dimensional plane, with the gateway at the origin. The $x$- and \mbox{$y$-coordinates} of each sensor are uniform random variables in the range \mbox{[30 m, 42 m]}. Each sensor transmits a measurement of size one byte every $t = 30$~s. The sensors can store at most \mbox{$b_{\max} = 10$} bytes of data. The maximum tolerable delay, defined as the time after which a measurement is no longer of interest to the gateway, is $d_{\max} \equals 3$ minutes. The relays are placed between the sensors and the gateway; their coordinates are chosen at random from the range \mbox{[10 m, 20 m]} while ensuring that any two relays are at least 1~m apart. The parameters $t_\mathrm{rx}$ and $t_\mathrm{tx}$ are $30$ s and $300$ ms, respectively. All links experience Nakagami-$m$ fading with $m \equals 1.2$ and a pathloss with exponent $\alpha=4$. The spreading factors are $s_\mathrm{sen} = 10$ and $s_\mathrm{rel} = 7$, respectively. Each transmission occupies a bandwidth of 125~kHz. A sender chooses one of three center frequencies (860~MHz, 864~MHz, and 868~MHz) uniformly at random for each frame. The transmit power is $14$\:dBm for all nodes. A channel code of rate $4/5$ is applied. A sensor ID is one byte in size. For this setup, a relay can include at most $v = 93$ measurements per frame.

The maximum number of past measurements a sensor can include in a frame is $r_{\max}$. This number should be chosen in a way that the sensor's storage capacity $b_{\max}$ is not exceeded, frames do not violate the duty-cycle constraint, and the oldest measurement is not older than $d_{\max}$. Thus, {$r_{\max} = \min \{b_{\max}, \hat{r}_{\max}, d_{\max} / t\}\:$} with \mbox{$\hat{r}_{\max} = \max\{r : f(r,s_\mathrm{sen}) \leq 0.01\}$}, which yields $r_{\max} = 6$ in our~setup.

This setup is representative of industrial use cases under investigation by the authors, where many sensors are installed over a small region with a harsh propagation environment. The high path loss results in much shorter communication ranges as compared to the kilometers-long ranges achieved outdoors.   Two metrics are used for performance evaluation (as in \cite{BBS19}): The \emph{measurement loss rate} (MLR) as the fraction of sensor measurements that the gateway fails to receive. The \emph{energy expenditure per delivered measurement} ($E_m$) as the average transmission energy that a sensor spends to successfully deliver a measurement to the gateway, obtained by dividing the energy spent by a sensor to transmit one frame by $1 - \mathrm{MLR}$.

Each simulation run corresponds to a system operation of three hours. The outcomes of multiple runs are averaged. The number of runs is such that at least 100 measurement losses are observed for each data point.

\begin{figure}
    \subfloat[Measurement loss rate.]{\label{MLR_vs_NumSensors}\includegraphics[scale=0.27,bb=-120 0 770 530]{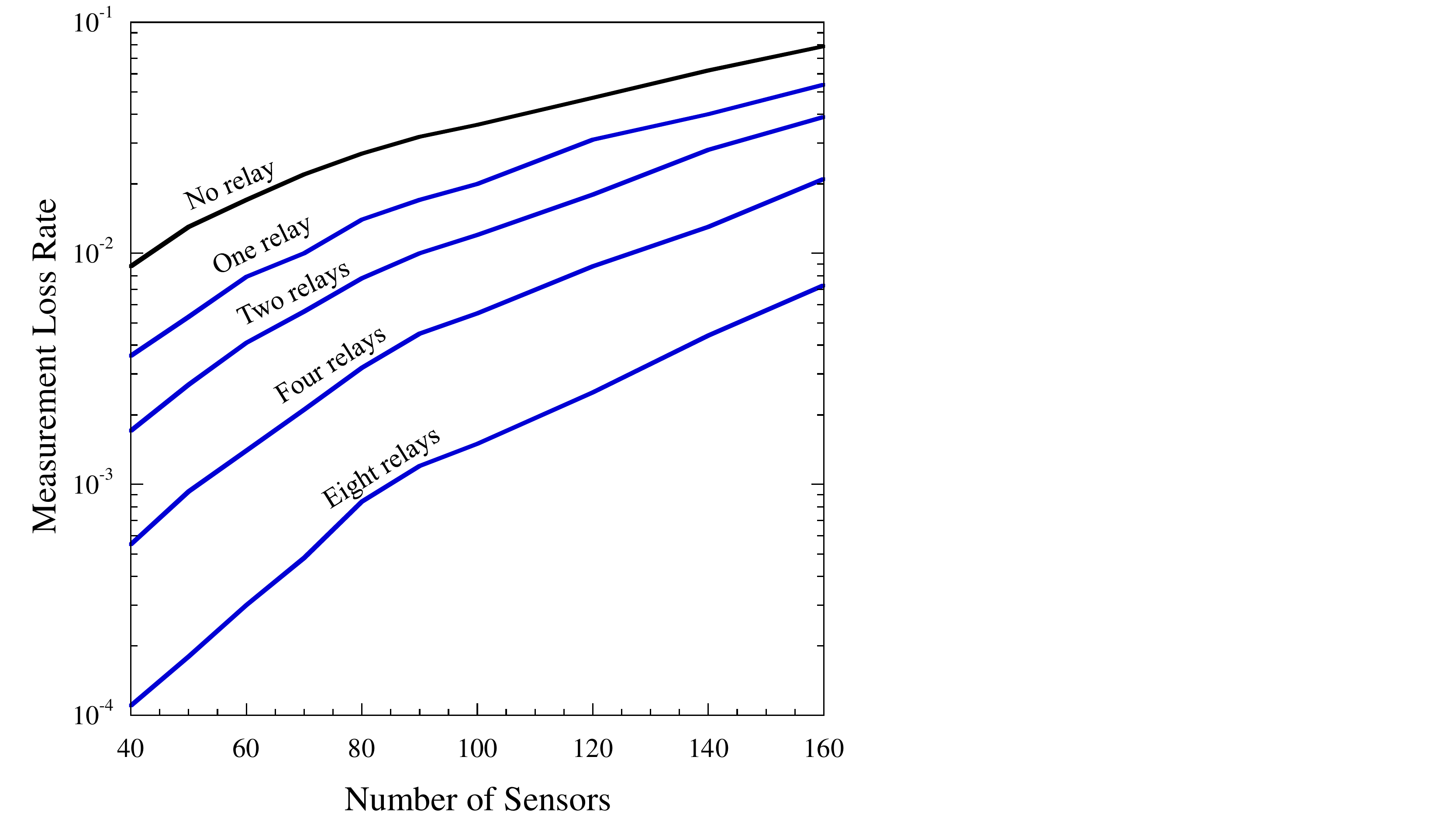}} \\
    \subfloat[Sensor energy expenditure per delivered measurement.]{\label{EDM_vs_NumSensors}\includegraphics[scale=0.275,bb=-120 0 770 520]{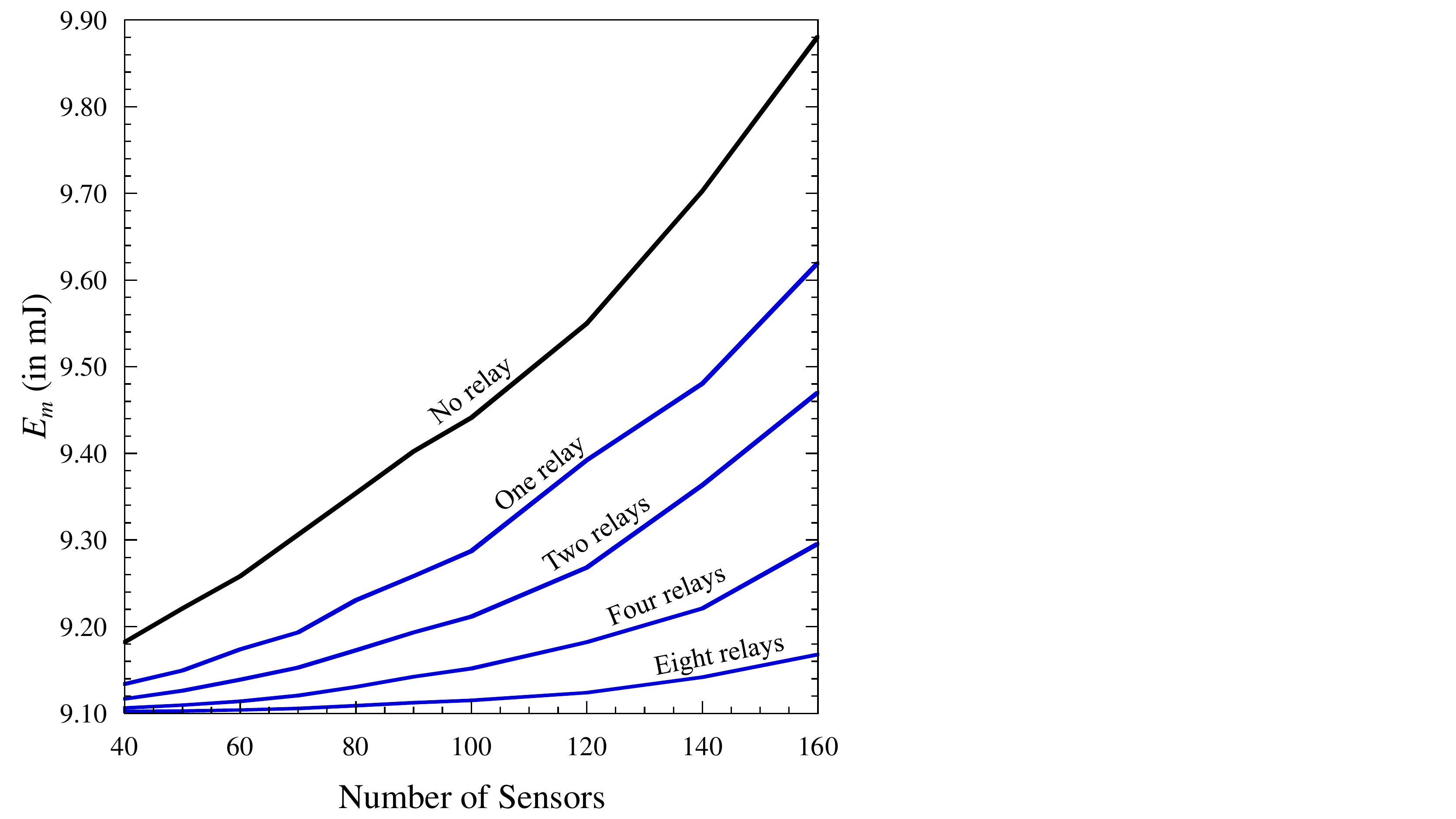}}
    \setlength{\belowcaptionskip}{-15pt}
    \caption{Measurement-delivery performance for $r = 3$.}
    \label{perf_vs_numsensors}
\end{figure}

\subsection{Simulation results}

Fig.~\ref{perf_vs_numsensors} shows the performance when sensors include four measurements in each frame ($r = 3$). Fig.~\ref{MLR_vs_NumSensors} demonstrates that the MLR increases with the number of sensors, as a result of rising interference. Relaying significantly improves the MLR: a single relay reduces the MLR by up to $50\,\%$; eight relays reduce it up to two orders of magnitude. Since a relay can forward at most $v = 93$ measurements at a time and \mbox{$\xi = t_{\mathrm{rx}}/t = 1$}, the probability of an overheard measurement being discarded by a relay is nonzero provided the number of sensors exceeds~93.  The results demonstrate that\,---\,even in the absence of coordination and taking into account the duty-cycle limit restricting the number of measurements that can be forwarded\,---\,relaying provides substantial benefits. A plot of the energy expenditure is shown in Fig.~\ref{EDM_vs_NumSensors}. Like the MLR, the $E_m$ also increases with the number of sensors and decreases with the number of relays.

\begin{figure}
    \subfloat[Measurement loss rate.]{\label{MLR_vs_NumRelays}\includegraphics[scale=0.31,bb=-120 0 770 530]{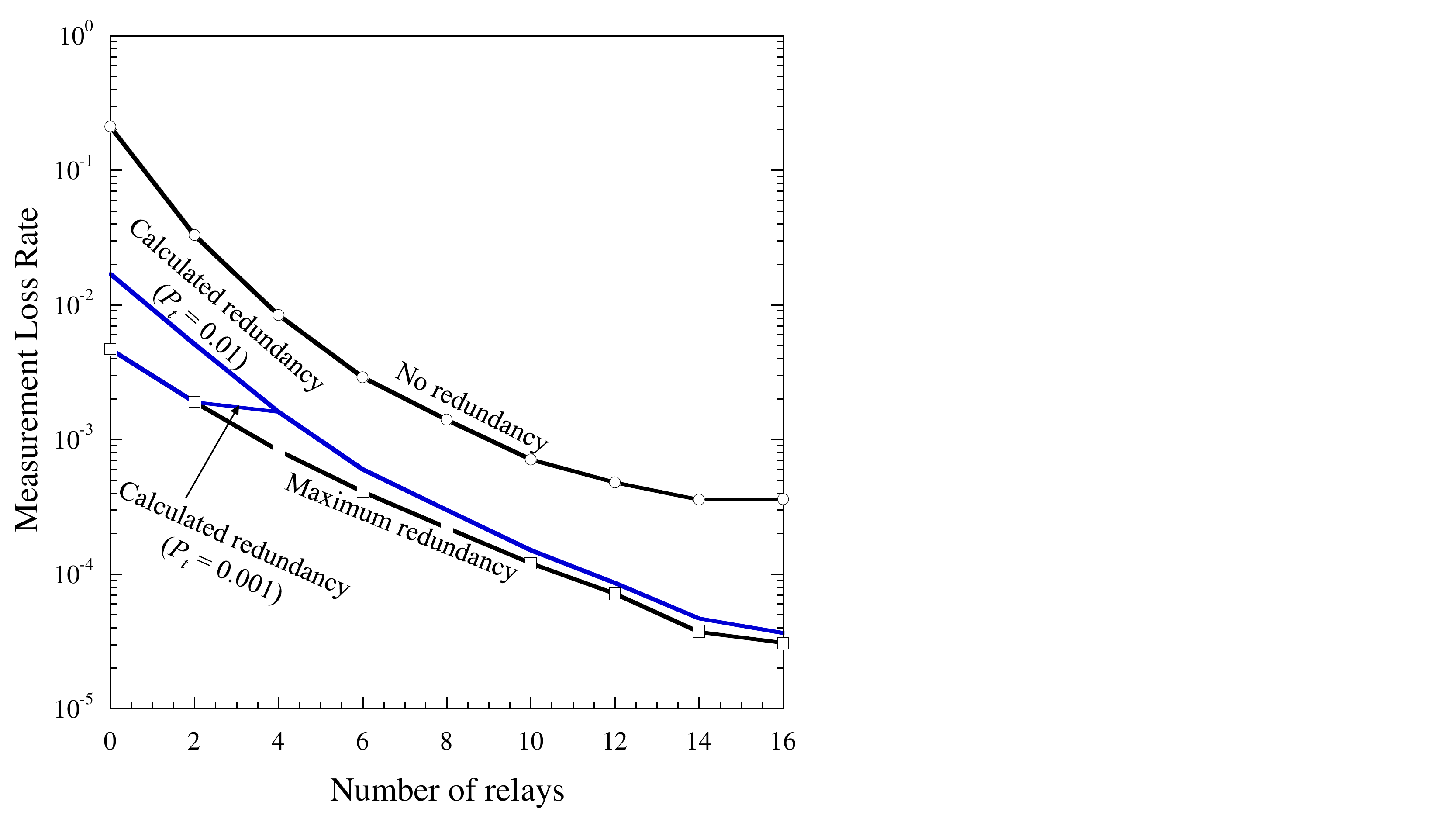}} \\
    \subfloat[Sensor energy expenditure per delivered measurement.]{\label{EDM_vs_NumRelays}\includegraphics[scale=0.31,bb=-120 0 770 520]{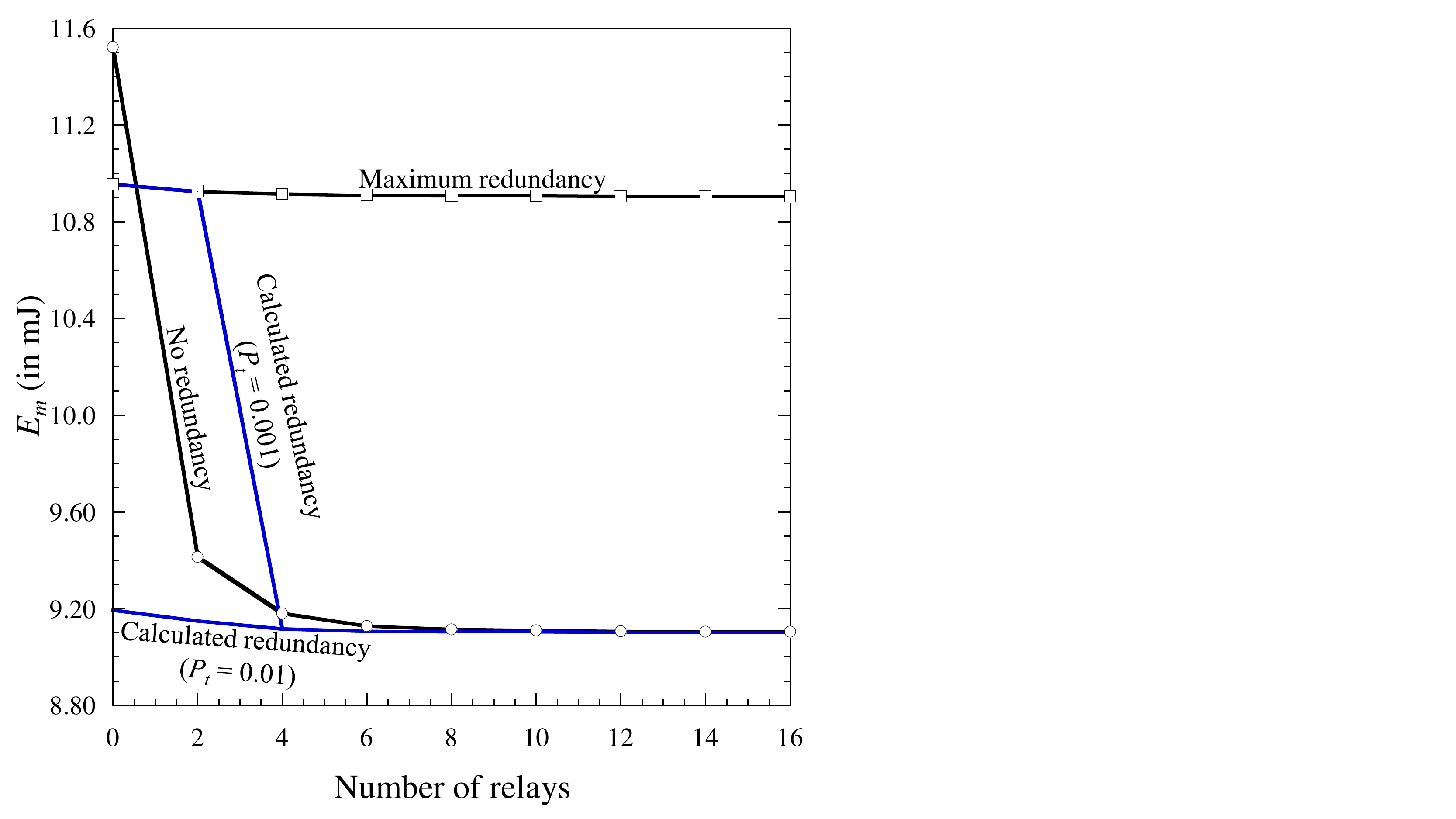}}
    \setlength{\belowcaptionskip}{-12pt}
    \caption{Measurement-delivery performance for $n = 60$.}
    \label{perf_res_n60}
\end{figure}

Fig.~\ref{perf_res_n60} compares the performance of three redundancy schemes in a network of  $n = 60$ sensors. One of the schemes employs no redundancy ($r = 0$, i.e., a frame contains only the current measurement). Another scheme sends the maximum possible redundancy per frame ($r_{\max}$). As demonstrated in Fig.~\ref{MLR_vs_NumRelays}, the transmission of redundancy reduces the MLR by up to two orders of magnitude. For both schemes, no redundancy and maximum redundancy, the MLR decreases with the number of relays.

Fig.~\ref{perf_res_n60} also shows results for a scheme in which the analysis of Section~\ref{reliability_analysis} is employed to calculate the number of past measurements to be included per frame. The procedure is an extension of the redundancy-allocation strategy of~\cite{BBS19}, into which we incorporate the effects of the relays. It is outlined in Algorithm~\ref{redundancy_algo}. The algorithm first determines the smallest value for the redundancy $r^* < r_{\max}$ for which $\mathrm{MLP}(r^*)$ is below a threshold $P_t$. If there is no such $r^*$, the value that minimizes the MLP is used. Because of LoRa's packet structure, multiple payload sizes may result in identical frame durations (and hence the same duty cycle and transmission energy). If $r^*$ is such that $\tilde r > r^*$ past measurements can be included in a frame without increasing the frame duration, we use $\tilde r$ to achieve higher reliability. The MLRs of this scheme are shown in Fig.~\ref{MLR_vs_NumRelays} for two values of the target MLP, namely $P_t = 0.01$ and $0.001$. Recall that the calculation of the MLP requires a probability distribution of the distances. In the evaluation of the MLP for Algorithm~\ref{redundancy_algo}, it is assumed that the distance of a sensor from the gateway is a uniform random variable in the range \mbox{[42 m, 59 m]}. While 42 m and 59 m are the minimum and maximum possible distances between the gateway and the sensors, the actual distances are not uniformly distributed between these values. Similar crude approximations for the other necessary distances are employed; as a result, the computation is not exact. Nevertheless, for $P_t = 0.01$, the MLR is below $P_t$ for most of the curve. For $P_t = 0.001$, we note that even the use of the maximum possible redundancy is not sufficient to ensure an MLR below $P_t$ with fewer than three relays. Two or fewer relays cause the algorithm to instruct the sensors to employ the maximum redundancy. Five or more relays allow the system to achieve an MLR below $P_t$.

The energy expenditures are shown in Fig.~\ref{EDM_vs_NumRelays}. While the maximum-redundancy scheme provides the best MLR, it also results in higher $E_m$ compared to other schemes; the only exception is the no-redundancy scheme without relays, which gives both highest MLR and highest $E_m$. The scheme that calculates the redundancy for $P_t = 0.01$ provides the best $E_m$ at the expense of a slightly worse MLR, thus achieving the best tradeoff between measurement loss and energy consumption.

\begin{algorithm}[t]
  \nl $\Omega = \{r: \mathrm{MLP}(r) \leq P_t, r \leq r_{\max}\}$\;
  \nl \uIf{$|\Omega| > 0$}{
    $r^* \gets  \min\{\Omega\}$\;
  }
  \Else{
    $r^* \gets \displaystyle \argmin_r \{\mathrm{MLP}(r): r \leq r_{\max}\}\:$ \;
  }
  \nl $\tilde r \gets \max \{r : t_f(r) = t_f(r^*), r \leq r_{\max}\}$
\caption{Procedure for redundancy allocation}
\label{redundancy_algo}
\end{algorithm}

\section{Conclusions and Outlook}
\label{conclusion}

In an industrial sensor network using LoRa to provide measurement readings to a gateway, the use of relays greatly improves communication reliability despite limited duty cycles and complete lack of coordination. Future plans include experimental evaluation of the relaying scheme.


\bibliographystyle{IEEEtran}
\balance
\bibliography{references}

\end{document}